\documentclass{elsarticle}
\usepackage{graphicx}
\usepackage{natbib}
\usepackage{color}
\begin{document}
\title{Limitations of the modelling of  crack propagating through heterogeneous material using a phase field approach}
\author{Herv\'e Henry}
\address{Laboratoire Physique de La Mati\`ere condens\'ee, \'Ecole
Polytechnique, CNRS, 91128 Palaiseau Cedex, France}
\begin{abstract}
The modeling of crack propagation in a heterogeneous material using a phase
field model is studied numerically in a simple test case: the crack meets a
wedge of higher fracture energy. It is shown that when the crack cannot enter
the wedge, phase field results are in qualitative agreement with theoretical
predictions with moderate  quantitative discrepancies. When the crack can
propagate in both regions, numerical results show that the interplay of diffuse interface
modelling used in the  phase field model with the interface between the two regions
induces spurious effects that are unphysical.
\end{abstract}

\maketitle
\section{Introduction}
   Heterogeneous materials are widespread in engineering and material science and they include 
  charged polymers, polymer blends after phase separation, concrete, composite materials, 
  alloys and biomaterials such as nacre or metamaterials \cite{Nacre,Composite1,Composite2}. 
  Their mechanical properties such as their elastic 
  coefficients can be computed using various strategies that rely on 
  the use of representative volume elements. However, the understanding of their
  resistance to failure is limited. 
  This is due to the fact that a crack propagating in a heterogeneous material
  under a simple loading will have a complex shape that cannot be predicted
  easily since it is the result of the  interplay of the fracture
  propagation laws and the microstructure of the material. When
  considering a heterogeneous  material, homogenization approaches that use
  experimentally determined quantities such as fracture energy can be made.
  However, one cannot predict crack properties for a given
  microstructure from the properties of the phases that compose the material. As
  a result,  designing the microtructure of the phases that compose a material
  to reach  given fracture energies\cite{Nacre,Composite1,Composite2} is
  impossible and a model that would allow such computation is highly desirable.

 In homogeneous material  the Linear Elastic Fracture Mechanics theory (LEFM)
 is efficient\cite{Freund90} when describing the propagation of a single crack in 2D materials. In the case of heterogeneous material,
 predictions made using the LEFM\cite{Hutchinson,Hutchinson2} are possible. But in
 general, the use of this  approach  is  difficult due to the  complex crack paths that
 are present. Moreover the extension of LEFM to homogeneous 3D systems is
 difficult. Therefore, in the case of 3D heterogeneous material, the application of LEFM theory  
 is likely to prove extremely difficult.  Among the different models of crack
 propagation, the phase field model could  describe correctly  failure of
 heterogeneous material because of its ability to describe complex
 crack paths in both 2D and 3D,.

Indeed, the phase field model of crack
propagation\cite{KKL,Francfort1998,Bourdin2008,Henry2004}, has been widely used to study crack
propagation in homogeneous material and  has been shown to be an effective way
of describing crack propagation in various contexts. It has been shown that
there is a very good quantitative agreement between the phase field model,
the LEFM and more importantly well controlled experiments 
in the case of complex crack paths such as 
the oscillating cracks in
a thermal gradient\cite{YuseSano,Ronsin95,Ronsin98,Corson2009} and the relaxation of crack fronts under mixed mode loading
(I+III)\cite{Ronsin2014,Pons2010,Henry2017,Lazarus2015,Leblond2011}.   
Since in the phase field model the fracture energy is a function of
the parameters of the coupled partial differential  equations it can
describe a crack propagating in a heterogeneous material where the
fracture energy  is varying slowly (on large lengthscales comapred to the process zone size)  by adding slow spatial variations  of the
model parameters. Such an approach has been used  to model  variations of the fracture energy anisotropy\cite{HakimKarma,Roman2018,Rethore2017} or of the fracture energy  \cite{Kuhn2016,Bourdin2014}.
It is then tempting to consider a crack propagating in a
heterogeneous material that consists of homogeneous regions with different
fracture energies that are separated by \textit{sharp} interfaces.

Some preliminary
simulations have shown results that are apparently qualitatively
correct\cite{Schneider2016,Bourdin2014}.
In the following I present numerical simulation
results that were performed in a simple test geometry. These results indicate
that in such a setup, the interface thickness between the two regions ($w_{int}$) has
dramatic effects on the numerical results unless it is small when compared to the phase field internal length ($w_\varphi$). Furthermore, even when the interface
thickness  seems to be small enough to ensure independence of the crack
propagation regime, results are unphysical. As a result, one should take extreme
care when considering crack propagation in heterogeneous material in the
presence of interfaces. The paper is organized as follows. First  the phase
field model is rapidly described and the mathematical description of the
heterogeneities is given. Thereafter, the numerical setup and the geometry that
has been studied  are presented.  This is followed by a description of the
numerical results. Finally the discussion  focuses on the questions that arise 
from the results and on a comparison with theoretical work relying on the LEFM theory.

\section{Phase Field Model and numerical implementation}
  While having distinct theoretical bases
   ranging from the  regularization of free
  dis\-con\-ti\-nuities\cite{Francfort1998} to a phenomenological approach\cite{KKL}
  inspired by  phase transitions, the so called phase field models of crack propagation  share the same
  basic idea. A free energy functional is written and the system evolves toward
  a minimum (local or global)  of this functional. The evolution toward the
  minimum can  be the result of a minimization of the free energy with respect to both
  the strain field and the phase field with, usually some constraints such as
  irreversibility. It can also be a relaxation equation for the phase field
  coupled to the wave equation (with possibly a dissipative term) for the strain
  field. In all cases it ensures the decrease of the free energy functional:
\begin{equation}\label{eq_F_g}
  \mathcal{F}=\int\left( \left(\frac{D}{2}|\nabla
  \varphi|^2+hV(\varphi)-\varepsilon_c
  g(\varphi)\right)+ g(\varphi)\left(\frac{\lambda}{2}(\mathrm{tr}\varepsilon)^2+\mu
  \mathrm{tr}(\varepsilon^2)\right)\right) \mathbf{dx}
\end{equation}
where $\lambda$ and $\mu$ are the Lam\'e coefficients and $\varepsilon$ is the
strain tensor.  The function that relates the phase field to the local elastic
moduli is $g$. Here, the phase field $\varphi$ goes from $1$
(undamaged) to $0$, totally broken and therefore $g$ ranges from 0 for
$\varphi=0$ to 1 for  $\varphi=1$. This choice corresponds to that  made in
\cite{KKL}. {It ensures, as was proven in \cite{KKL} that in the limit of vanishing interface thickness, the equilibrium solution converges toward a traction free crack.} The potential 
$h V-\varepsilon_c  g$  can either be a tilted double well
\cite{KKL} (tilted by $\varepsilon_c$) or a single well
potential\cite{Francfort1998}. It can also be a single obstacle potential, as in
\cite{Schneider2016}. All these choices lead to the same qualitatively correct  behaviour
in 2D and yield very similar results when describing the propagation of a single
crack\cite{Kuhn2015}. Here, following\cite{KKL}, the coupling function $g$  has been chosen to be
\begin{equation}
  g(\varphi)=4\varphi^3-3\varphi^4,
\end{equation}
so that the 1D equilibrium crack solution corresponds to a situation where the
residual stress in the unbroken material vanishes. The potential
$hV$ is  equal to $h(1-\varphi)^2\varphi^2$.{ With such a choice,
the \textit{actual} potential in the  model of \cite{KKL} is
$hV(\varphi)-\varepsilon_c g(\varphi)$, which , since the coupling function has
a zero slope  in $\varphi=1$ leads to a purely elastic unbroken
phase\cite{Kuhn2015} similarly to what is observed when considering single
obstacle type models\cite{Schneider2016,Tanne2018,SARGADO2018}.{ Here we have chosen $h=1$, $D=1$ and $\varepsilon_c=1 $ }.

An important  difference between the single obstacle model and the double well
model is the way crack  nucleation occurs.  This will be qualitatively  discussed now.
In the single obstacle model\cite{Tanne2018,Kuhn2015}, the nucleation occurs as
soon as the load is such that the elastic energy density is above the
\textit{propagation threshold} (which, with notations used here,  is equal to
$\varepsilon_c $ if interface curvature effects are neglected.)  in a large
enough region (where large enough is above a few material lengthscales $
D/(h,\varepsilon_c) $ ). In the double well model, the elastic energy density
should be above another threshold (that is larger by $\approx h/\varepsilon_c $)
in a large enough region. Hence, in the single obstacle model a crack is likely
to  nucleate in a domain where the elastic energy density is below the
propagation threshold almost everywhere. On the other hand, in the double well
potential when a crack nucleates, it is likely that the elastic energy density
is above the propagation threshold in a macroscopic domain around the nucleation
site\footnote{For a discussion on a system similar to cracks see
\cite{Corson2010}}. Therefore  the two models will probably exhibit dramatically
different post nucleation behaviour. In the same spirit, the  behaviour when the
crack propagates under extremely high loads (close to material strength) should
be different. Here, since nucleation is not studied this difference between the
models can be safely ignored. In the same spirit, the fact that nucleation is
not considered here, implies that the description of the crack propagation by
the phase field model is correct when the phase field length scale $w_\varphi$
is much smaller than the macroscopic lengthscale, and when the phase field model
parameters give the correct fracture energy. This is indeed confirmed by
simulations performed (not shown)  using the single obstacle model used in
\cite{Tanne2018}. }
 
 The
 evolution equations for the phase field and the displacement field are:
 \begin{eqnarray}
   \partial_t \varphi&=&-\beta p\left(\mathrm{tr} \varepsilon,\frac{\delta \mathcal{F}}{\delta
   \varphi} \right)  \left(\frac{\delta \mathcal{F}}{\delta
   \varphi}\right)  \label{eqdtphi}\\
   &=&-\beta p\left(\mathrm{tr} \varepsilon,\frac{\delta \mathcal{F}}{\delta
   \varphi} \right) \left( D\Delta \varphi -hV'(\varphi)+\varepsilon_c g'(\varphi) -g'(\varphi)\left(\frac{\lambda}{2}(\mathrm{tr}\varepsilon)^2+\mu
  \mathrm{tr}(\varepsilon^2)\right)\right)    \\
   \rho \partial_{tt} u_i&=&-\frac{\delta \mathcal{F}}{\delta
   u_i}+D_{diss}\Delta \dot{u_i}\label{eqdtu}
\end{eqnarray}
where $\rho$ is the mass density equal to 1,{\color{red}{ $\beta p$ is a kinetic parameter that has three purposes:
\begin{itemize}
  \item  $\beta$ is proportional to the energy dissipation in the process zone  
\cite{Henry2008}.
  \item $p$ ensures crack irreversibility: if $$\frac{\delta \mathcal{F}}{\delta
   \varphi}<0$$ then $p=0$
 \item  $p$ distinguishes between compression and extension if  $\frac{\delta \mathcal{F}}{\delta
   \varphi}>0$: if $\mathrm{tr} \varepsilon>0$ (i.e. under extension) then $p=1$. On the contrary  if $\mathrm{tr} \varepsilon<0$  (i.e. under local compression),  $p$ is a \textit{positive}\footnote{The fact that $p$ is positive ensures the decrease of $\mathcal{F}$. } number chosen so that $\partial_t\varphi$ is the minimum of $0$ (to keep irreversibility and avoid crack healing under compression) and
   $$
   -\beta  \left( D\Delta \varphi -hV'(\varphi)+\varepsilon_c g'(\varphi) -g'(\varphi)\left(\frac{\lambda-K_{\mathrm{Lame}}}{2}(\mathrm{tr}\varepsilon)^2+\mu
  \mathrm{tr}(\varepsilon^2) \right)\right)  
   $$
   where $K_{Lame}=\lambda+2\mu/3$ is the bulk modulus.
\end{itemize}
While the first rule ensures energy dissipation at the crack tip, the two last one simply ensure crack irreversibility and  the fact that energy due to compression does not participate to crack advance.  
}}

\begin{figure}
  \centerline{\includegraphics[width=0.6\textwidth]{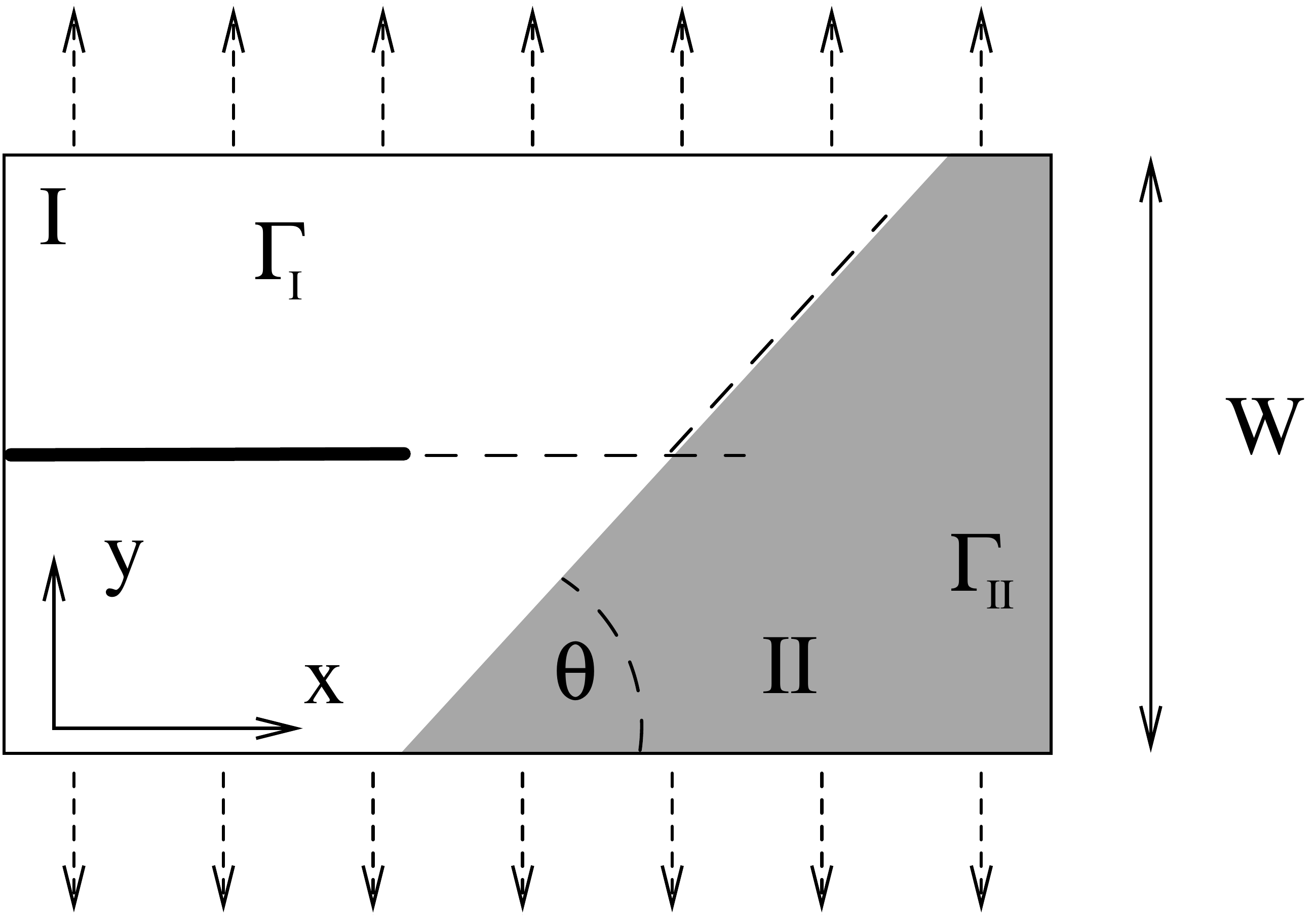}}
  \caption{\label{figsetup} schematic view of the material, the initial crack
  and the coordinates.The fracture energies in white (resp. grey) regions are
  $\Gamma_I$ (resp. $\Gamma_{II}$). The grey region is called the obstacle in the manuscript.{\color{red}The elastic boundary conditions are imposed displacement $u_x=0$ and $u_y=\pm \Delta_y/2$ at $Y=\mp W/2$  and at $x=0$ and $x=L$ $u_x=0$ and $u_y=\Delta_y (y/W)$}}
\end{figure}

$D_{diss}$ is a \textit{fluid viscosity}. Two
limiting cases for $D_{diss}$ have been used here. In the first  $D_{diss}=0.01$   is
taken small to ensure a negligible  dissipation during wave propagation.
{ The actual dynamic crack propagation is modeled with the possibility of branching events at high crack speed.\cite{Karma2004,Henry2008,bleyer2017,Henry2014}}. In the  second it is chosen so that the  relaxation toward mechanical equilibrium of the elastic field is fast so that  the system mimics well quasistatic crack propagation ($D_{diss}=0.4$)\footnote{When $D_{diss}$ is large  relaxation is
 very slow since dissipation prevents a rapid motion of the material point, when
 $D_{diss}$ is small the oscillations are undamped and relaxation toward
 mechanical equilibrium is slow.\cite{Henry2017}. The value used here was determined by minimizing  the relaxation time of an elastic sheet with a non moving crack toward equilibrium }. 
 Here, it has been found that for a given load and fracture energy, the qualitative behaviour of crack (i.e. whether they propagate or not) along the interface with the obstacle (see fig. \ref{figsetup}) is independent of the value of $D_{diss}$ in these two limiting cases.

This work is limited to the study of a crack interacting with a single
inhomogeneity or obstacle  which is characterized by a higher fracture energy than the matrix, while it has the same elastic coefficients. 
Such a choice is clearly unrealistic. However the purpose of
this paper is to show the issues that arise from introducing interfaces between
regions of different fracture energies  in the phase field model. Therefore
in order to avoid the additional complexity  that will  arise from
heterogeneities in the elastic coefficients, 
they are taken equal in both phases. In the same spirit of simplification a pure mode I crack
propagating in a homogeneous material until it reaches a flat interface with a
region of higher fracture energy is considered as it can be seen in fig.
\ref{figsetup}.{  In the early work of Karma and
coworkers\cite{KKL}, it was shown that in the phase field model, the fracture
energy is: 
\begin{equation}
  \Gamma=\int_0^1 \left(\sqrt{D(\varepsilon_c(1-g(\varphi))+hV(\varphi) )}
  \right)d\varphi,
\end{equation}
and  is a function of $D>0$, $\varepsilon_c>0$ and $h\geq0$. In 1D, the exact
shape of the phase field profile and the material lengthscale are also a
function of these parameters. The exact interface  profile shape (up to a
rescaling by the material lengthscale)  is a function of the respective
magnitude of  $h$ and $\varepsilon_c$ while for a given ratio $h/\varepsilon_c$,
the material lengthscale $w_\varphi$ is proportional to
$\sqrt{D/\varepsilon_c}$. Since this work is solely focused on the propagation of a crack along an obstacle  it has been chosen to avoid  the additional complexity of describing varying material lengthscales and interface profiles. To this purpose the free energy functional, following \cite{KKL} is  chosen:}
\begin{eqnarray}\label{eq_F}
  \mathcal{F}&=&\int \left((1+k(\mathbf{r}))\left(\frac{D}{2}|\nabla
  \varphi|^2+hV(\varphi)-\varepsilon_c
  g(\gamma)\right)+\right .\nonumber \\ 
  && \left. g(\varphi)\left(\frac{\lambda}{2}(\mathrm{tr}\varepsilon)^2+\mu
  \mathrm{tr}(\varepsilon^2)\right)\right)\mathbf{dx}
\end{eqnarray}
With such an expression, the shape of the equilibrium solution for the fracture
is independent of $k(\mathbf{r})$ and the fracture thickness $w_\varphi$ is kept constant
while the local fracture energy of the simulated material is $\Gamma_{II}=\Gamma_{I}
(1+k(\mathbf{r}))$.  { As a result, the material strength in the
region II is  $(1+k(\mathbf{r}))$ times higher than in the region
I\cite{Tanne2018}. However, here, $w_\varphi$ is solely a regularization
lengthscale and can be taken to be  large when compared to actual material
lengthscales as long as the following conditions are met: it is much smaller than macroscopic lengthscales and  nucleation is not present.}

Here, $k(\mathbf{r})$ is, using Cartesian coordinates
\begin{eqnarray}
  k(x,y)&=&a_0(1+\tanh(d/w_{int}))/2 \mbox{ with }\\
  d&=&\cos\theta (x-x_{off}) + \sin \theta (y)
\end{eqnarray}
where $a_0>0$ is the amplitude of fracture energy increase, $d$ is the signed
distance between  a point $(x,\  y)$  and a straight line making an
angle $\theta$ with the direction of propagation of the crack (see fig.
\ref{figsetup}).  $w_{int}$ is the  model parameter that measures the interface
thickness between the regions I and II. For instance with  $a_0=1$, the fracture energy is doubled
when crossing the interface. In this context the crack is expected to either
propagate along the interface,  to stop or to propagate through the tougher
region II. In the following we will present numerical results obtained with a
phase field model of crack propagation in this setup and discuss whether they
correspond to expected behaviours, keeping in mind that here the interface is a
pure geometrical object that has no specific properties such as interface
toughness. 

These PDEs (\ref{eqdtphi}, \ref{eqdtu}) were simulated, using a
finite difference scheme with a simple forward Euler time stepping scheme on a
rectangular grid (see fig.\ref{figsetup}). The
grid spacing ($\delta x=0.3$)  was chosen so that  the phase field interface thickness (i.e. half the crack
thickness) is approximately 4 grid points which ensures that lattice pinning can
be neglected.{ In addition, simulations using a coarser grid and a finer grid ($\delta x=0.3/\sqrt{2}$ and $\delta x=0.3\sqrt{2}$) gave the same results with an error smaller than a few percents}.
The boundary conditions for the elastic field are fixed
displacements:
\begin{equation}
  u_y= \pm \Delta y/2,\ \  u_x=0  \mbox{ at } y=\pm W/2
\end{equation}
 The  elastic energy  stored in the material is then 
 $(\lambda/2+\mu)  \Delta y^2/W$ where $\lambda$ and $\mu$ are the Lam\'e
 coefficients. The system size $W$ was varied between   216 and 864  without any
 quantitative changes, when the elastic energy stored in the strained material 
 was kept constant. { This indicates that the ratio between interface thickness, the so called the internal length, of the phase field model and the system size does not affect the results. Therefore, it can be safely assumed that the simulatoions were performed in a limit for which the internal lengthscale is small enough to ensure that the model approximates well the  infinitely sharp crack of the LEFM.}

\section{Results}

\subsection{Infinitely tough region}
\begin{figure} 
  \centerline{\includegraphics[width=0.6\textwidth]{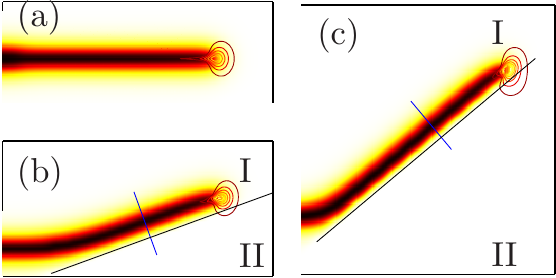}}
   \caption{\label{figcontoursinf}Contour plot of the elastic energy density at
   the end of a simulation for which the crack has stopped superimposed on the phase field color plot (white corresponds to $\varphi=1$ and black to $\varphi=0$. The straight black  line
   corresponds to the boundary of the obstacle.
   The plots (a) corresponds to a freely propagating
   crack.  Plot  (b)  to a stopped crack with  $\theta=20^o$ and (c) to a
   stopped crack with  $\theta=40^o$. It is worth noting that in the later case,
   the maximum of the elastic energy density is higher than in the case of a
   freely propagating crack.
   {It is worth mentioning that the contour plot of the elastic energy density present a tail in the fully broken region that is due to the slow relaxation of the phase field toward its zero equilibrium value in the broken region. The phase field profile and along the blue lines are plotted in fig.\ref{figprofing}.}}
 \end{figure}

 We first describe the propagation of a crack that meets an infinitely tough
 region where the phase field is set to 1 and does not change with time.
 In this case, the crack can only  propagate along the interface  between the
 normal medium and the infinitely tough medium. 
 The questions that have  to be answered  are whether the crack
 propagates and at what velocity. Here we will mostly  limit ourselves to the first
 since a theoretical prediction can be made. 

  To this end, we 
  neglect the effect of the crack propagation velocity and suppose 
 that the stress intensity factors before the kink are 
 \begin{equation}
   (K_1,K_2)=(A \Delta_y,0)\label{eq_Kstraight}
 \end{equation}
 where A is a  real constant that depends on the geometry of the material, and
 $\Delta_y$ is the applied displacement at the boundaries of the system. From
 this, using equations (2), (3) and (66) from \cite{Amestoy1992}, the stress
 intensity factors after deflection $(K_1(\theta,K_0),\ K_2(\theta,K_0))$ can be
 computed as functions of the kink angle $\theta$ and of the stress intensity
 factor before the kink $K_0$. In our particular case where $K_2=0$ before the
 kink the SIFs after the kink  write, using the notation of  \cite{Amestoy1992} at zeroth  order in
 the length of the kinked crack~:
 \begin{equation}
  K_1=F_{11}(\theta) K_0,\ K_2=F_{21}(\theta) K_0 \label{eq_Kkink}
 \end{equation}
 where $F_{11} $ and $F_{21}$ are universal functions of $\theta$.  From this 
 the energy release rate associated with the crack propagation can be computed:
 \begin{equation}
   G(K_0,\theta)=(K_1(\theta,K_0))^2+(K_2(\theta,K_0))^2. 
 \end{equation}
 One expects the crack to propagate when $G(K_0,\theta)$ is larger than the fracture
 energy $\Gamma$ (which can be computed using the parameters of the phase field model\cite{KKL}). 
 
  In order to have an estimate of $\Gamma$ that takes into account the possible
  errors induced by   the numerical implementation,  we  compute the
 threshold value $\Delta_{yc}$ above which the straight crack starts to
 propagate, and from this we have: 
 $\Gamma=(A\Delta_{yc})^2$.  As a result, one expect the crack to propagate as
 soon as $(K_1(\theta,K_0))^2+(K_2(\theta,K_0))^2$ is larger than
 $(A\Delta_{yc})^2$. Taking into account equations
 (\ref{eq_Kstraight}, \ref{eq_Kkink}), this translates into
 \begin{equation}
   \Delta_y^2\times(F_{11}(\theta)^2+F_{21}(\theta)^2)>
   \Delta_{yc}^2\label{threshold_kink},
 \end{equation}
 where the proportionality constant $A$ is no longer present. 

 \begin{figure}
   \centerline{\includegraphics[width=0.6\textwidth]{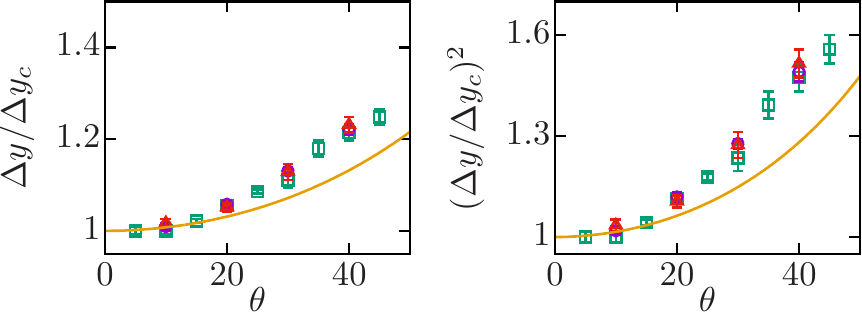}}
   \caption{\label{plot_inf_th}\textbf{Left}: Plot of the theoretical
   prediction of the threshold of propagation (solid line) together with the
   propagation threshold obtained from simulations. The system size was $W= W_0=
   $ (square),   $W= 2W_0$ (circles) and  $W= 4W_0$ (triangles). For all system
   sizes, $\Delta_y$ has been rescaled by the threshold value for propagation
   $\Delta y_c$ and error bars are given. \textbf{Right:} same plot using the
   fracture energy (i.e. $(\Delta y/\Delta y_c)^2$ instead of the applied
   displacement. }
 \end{figure}

 In the situation described above the crack is forced either to kink or
 to stop propagating, so we expect that the 
 threshold $\Delta_y$ above which the crack propagates along the interface to
 obbey the eq. (\ref{threshold_kink}). The result of simulations are now described
 and compared to the prediction of eq. (\ref{threshold_kink}).
  
 As expected,  two possible behaviours are  observed~: either the
 crack stops or it propagates along the interface. When the crack stops, due to
 inertial effects,  it propagates over a small
 distance along the interface before stopping. In figure \ref{figcontoursinf},
 the crack interface profile corresponding to $\varphi=0.5$ is plotted together
 with the obstacle and the elastic energy density in the three cases (crack
 propagating freely, crack propagating along the interface and crack stopped
 after a short transient). In all three cases, the applied load $\Delta_y$ is the
 same. 
 
  The duration  of the transient regime before the crack reaches a steady
 state (either stopped or propagating along the interface) and  can be tuned by
 changing the amplitude of the dissipation terms. However the nature of the steady state
 itself is independent of that  dissipation terms, and a comparison
 between theoretical prediction and numerical results can be made.

  In figure \ref{plot_inf_th} the prediction is plotted  together with the
 results of the phase field simulations for angles ranging from $0^o$ to $45^o$
 and three distinct system sizes, the phase field model parameters, being kept
 unchanged.  First one must note that the simulations with different system size
 give very similar results indicating that the system size is large enough
 compared to the phase field model interface thickness to ensure a good
 convergence of the model. The invariance with respect to the system size also
 indicates that the zeroth order expression of ref.\cite{Amestoy1992} is
 sufficient to describe the perturbation of the stress intensity factors at the
 crack tip due to the kink.  When turning to the results strictly speaking it is
 obvious that  the trend of the phase field data is very similar to the one of
 the analytical prediction with a  rather small discrepancy at low kink angles.
 The phase field data  slightly  overestimates the threshold value.
 Nevertheless,  there is a significant overestimation of the increase in
 \textit{fracture energy} induced by the kink, especially at angles above
 $30^o$.
 \begin{figure}
   \centerline{\includegraphics[width=0.6\textwidth]{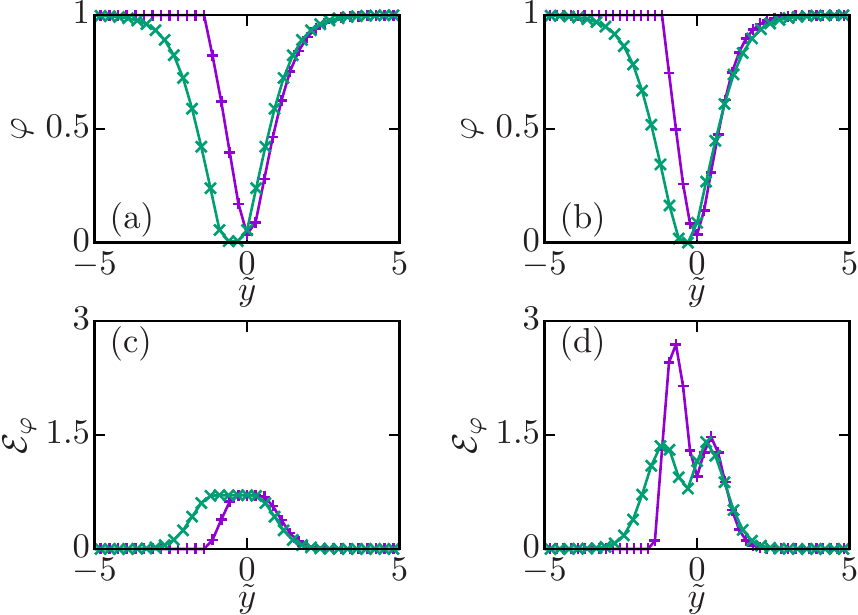}}
  \caption{Solid lines: phase field profile perpendicular to the crack path before (green X) and after (purple +)  meeting the obstacle  for $\theta=20^o$(a) and $\theta=40^o$(b). The distance from the crack midline is denoted $\tilde{y}$. The \textit{phase field} energy density is plotted in (c) and (d). The line along which the profiles are plotted are represented in fig. \ref{figcontoursinf}. \label{figprofing} }
 \end{figure}

  Plotting the phase field profile  in fig.\ref{figprofing} gives a first hint
  of the origin of  this discrepancy. One can see that on the obstacle
  side, the phase field profile is significantly steeper~: the exponential tail
  is no longer present and in the interface region  the profile is also much
  steeper. This translates into the fact that the fracture energy along the
  obstacle is  higher than  the fracture energy in the free propagation region. A
  numerical calculation  of the fracture energy \footnote{The fracture energy
  is  estimated as  $\cos(\theta) \int_{-W/2}^{W/2} dy
  \frac{D}{2}|\nabla_{2D}\varphi|^2+hV(\varphi)-\varepsilon_c(g(\varphi)-1) $ in
  the region where the crack is supposed to follow the interface with the
  obstacle. The same formula with $\theta=0$ is used in the free propagation
  region.} shows that there is a relative  increase in it   of
  approximately of  $6 $  (resp. 17)  percent for $\theta=20^o$ (resp $40^o$).
  Such values are consistent with the deviation from the prediction of
  eq. (\ref{threshold_kink}) which is also  of 5 (resp. 16 ) percent.  
  { The fact
   that, as in \cite{SARGADO2018} the crack profile is steeper close to the obstacle is clearly  related to
  the change in the  elastic energy density field at the crack tip induced by the crack
  kink}. Indeed, when plotting the phase field contour line for $\varphi=0.5$,
    that is  a good approximation of the crack profile together with the contour
  lines of the  elastic energy density (multiplied by $g(\varphi)$), one can see
  in fig.\ref{figcontoursinf}, that the maximum elastic energy density at the
  tip of arrested cracks is higher than at the tip of a freely propagating
  crack. It is also located closer to the boundary of the obstacle. As a result the
  local driving force toward $\varphi=0$ is higher on the obstacle side. This
  leads to a steeper profile (and an higher fracture energy). Hence the mode
  mixity at the crack tip induced by the imposed kink induces a change of the
  fracture profile,  and as a consequence an unphysical increase of the fracture
  energy. In the well documented case $\theta=90^o$\cite{Hutchinson,Hutchinson2,Moes2011}, simulation results (now
  shown) indicate that the crack will propagate along the interface for $G>8G_c$.
  This is much  higher than the predicted threshold of $4G_c$ and is expected since
  the effects of the crack profile steepenning are higher.

  Here we have seen that when encountering an infinitely  tough region the phase
  field crack behaviour is qualitatively correct. However  some discrepancies
  with theoretical predictions are present due to a change in the fracture
  profile coming from the presence of the infinitely tough region. This gives
  confidence that the phase field model will be  able to describe qualitatively
  well the crack front path in a complex heterogeneous material where one phase
  cannot be broken.
   \subsection{Two regions of finite toughness}
 \begin{figure}
   \centerline{\includegraphics[width=0.6\textwidth]{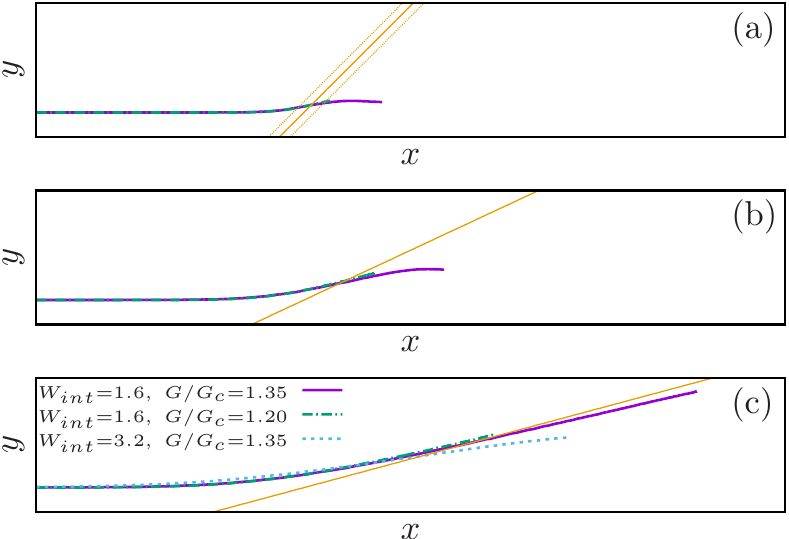}}
    \caption{\label{figthickint} (a), (b), (c) Trajectories of the crack tip for 
    $\theta=45, 25 \mbox{ and } 15$ degrees respectively.  The thick orange line corresponds to the interface between the region I and II. In (a) two dotted orange lines are plotted 1.6 space units  appart from the interface. 
    The correspondence between line style and parameter values is given in (c).  It is worth mentioning that the load is such that it corresponds to crack propagating along the interface in fig. \ref{plot_inf_th}.}
  \end{figure}

  This section describes what happens  when a crack meets a region
  of finite toughness and more specifically it 
  focuses on the effects of the choice of the mathematical description of the
  interface between the two regions. 
  We  consider a 
   situation where the  values of the load, and the fracture energy
  in both regions are such that the crack cannot propagate steadily in the
  region II. The results presented here have been 
  obtained with $\Gamma_{II}=2\Gamma_I$, but the same results were obtained with higher ratios $\Gamma_{II}/\Gamma_I$.
  The interface is described as a region of finite thickness $w_{int}$ over
  which the fracture energy goes from $\Gamma_I$ to $\Gamma_{II}$:
  \begin{eqnarray}
    \Gamma(x,y)&=&\Gamma_{I}+(1-\Gamma_{II})(1+\tanh(d/w_{int}))/2,\\
     d&=&\cos\theta (x-x_{off}) + \sin \theta (y).
  \end{eqnarray}
   If the mathematical description of the interface between the two regions is
  correct, the crack propagation regime should 
  be the same as in the previous section and the results should agree (at least
  qualitatively well) with the analytical  prediction of eq.
  (\ref{threshold_kink}). In this framework, there are three control parameters:
  the angle $\theta$  and the applied load (that were already present in the
  previous section) and the interface thickness between the two phases that will
  be denoted $w_{int}$. The first two parameters are physically relevant  and
  the behaviour of the crack should solely be a function of them while the latter
  is a model parameter that must  be chosen properly to ensure that the crack
  motion is correct.  

  Before turning to the results, it is useful do discuss the the values of $w_{int}$ for which the model
  results  can be expected to be correct. In the situations where
  the fracture energy is smoothly varying over  lengthscales that are much
  larger than the fracture interface thickness, it has been shown that 
  \cite{HakimKarma,Kuhn2016} the phase field model is able to describe the crack path in
  good agreement with LEFM theory. In this situation, the \textit{force} that pushes
  the crack path toward a kink is proportional to the fracture energy gradient
  while the force that pulls it back toward a straight path is proportional to
  the kink angle. As a result 
  when $w_{int}>>w_\varphi$, the crack will, probably, move inside the
  boundary between the two domains with, possibly, a slight kink that will be a
  function of both  $\theta$ and  $w_{int}$ and it will eventually
  stop. This behaviour differs dramatically from what is predicted and observed 
  when a crack meets a region of high toughness.
   Therefore, it is
  expected that the proper behaviour  is observed when $w_{int}$ is small enough. In the following
  the reality of these expectations will be compared to actual numerical results
  and the implication of these results will be discussed.

  \begin{figure}
    \centerline{ \includegraphics[width=0.45\textwidth]{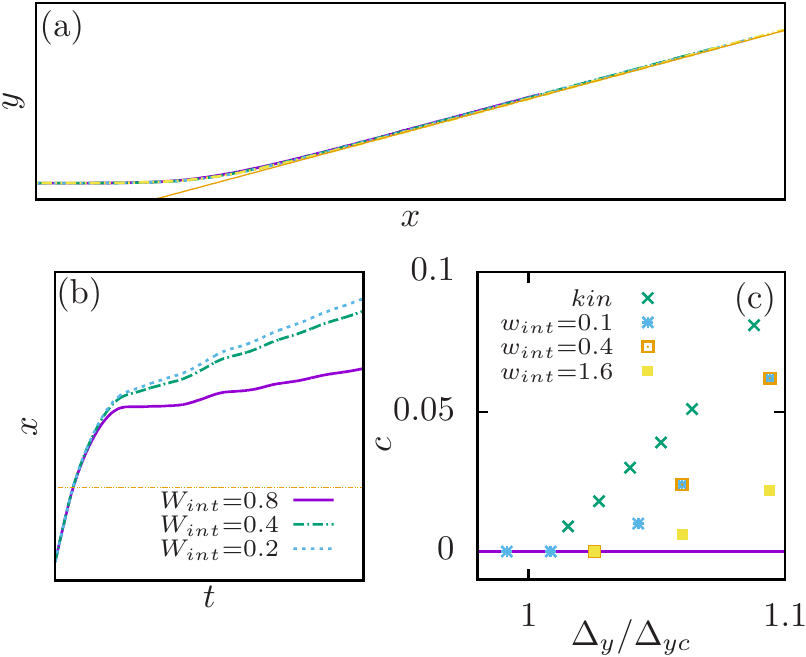}}

    \caption{\label{figthinwintlowtheta}$\theta=15^o$ (a)Plot of the crack tip trajectories
    for different values of the interface thickness and a same value of load.
    (b) Plot of the $x$ position of the crack tip as a function of time for
    different  different values of the interface thickness. (c) Plot of the crack
    velocity as a function of load for different values of the interface
    thickness. In (c) $\theta=10^o$.}
  \end{figure}

  To this end  we
  first consider the large $w_{int}$ limit. In this case, as expected and for
  all values of the tilt angle taken between 10 and 45 degrees the crack, when
  hitting the interface keeps on propagating until the \textit{local} fracture
  energy is large enough to prevent crack propagation. Typical illustrations of
  this behaviour are shown fig (\ref{figthickint}) for angles 15, 25  and 30 degrees 
  and different values of the applied load and two different values of
  $w_{int}$ that are  of the same order of magnitude as the fracture interface
  thickness\footnote{$w_{int}$ is  1.6 or 3.2 grid points while the fracture
  interface thickness is about 4 to 5 grid points as can be seen in
  fig.\ref{figprofing}}. In all cases,   the crack behaves as it were propagating in a smooth fracture energy
  landscape, which  does not correspond to the behaviour expected
  when the crack is meeting an interface between two regions with different
  fracture energies. 

  We now turn to the case where $w_{int}$ is small. In this case, a deflection of
  the crack path approximately along the interface between the two regions is
  expected. In a first set of simulations with small values of $\theta\approx
  10^o$ it is shown that this behaviour can be actually observed and the
  convergence of some measurable  quantities is discussed. To this end  we
  consider first a load value for which a crack was propagating along the
  interface in the infinitely tough region II case and consider different values
  of $w_{int}$. As can be seen in fig. \ref{figthinwintlowtheta} (a), for
  sufficiently small values of $w_{int}$ the crack path follows very well the
  interface  and the trajectories are nearly indistinguishable. Nevertheless,
  when considering the position of the crack tip as a function of time, there is
  a marked difference between the two values of the crack velocity when it is
  travelling along the interface as can be seen in fig.
  \ref{figthinwintlowtheta} (b).  In fig. \ref{figthinwintlowtheta} (b) the
  results are summarized:  the crack velocity is
  plotted against the applied load for different values of $w_{int}$. The plot  indicates
  that there is a convergence toward a limiting value of the crack velocity when
  $w_{int}$ goes to zero.  This limiting value is smaller than the one
  obtained in the kinetic case. This translates  into a higher threshold for
  crack propagation along the interface and, as a result, into a higher
  discrepancy with the analytical predictions. Hence, when considering small
  values of $w_{int}$ and small tilt angles, the phase field model is able to
  reproduce qualitatively the results obtained using the LEFM. Nevertheless
  there is a significant quantitative discrepancy which  is clearly
  related to the fact that the crack (defined as the region where $\varphi$
  differs from 1) has entered the high energy region over at least one grid
  point due to the mode mixity that favours the propagation into the high
  fracture energy region. This results in an increase in the apparent fracture
  energy that is higher than the one due to the steepening of the $\varphi$
  profile that was induced by the effects of  both  mode mixity and kinetic
  blockage.

  \begin{figure}
    \centerline{\includegraphics[width=0.6\textwidth]{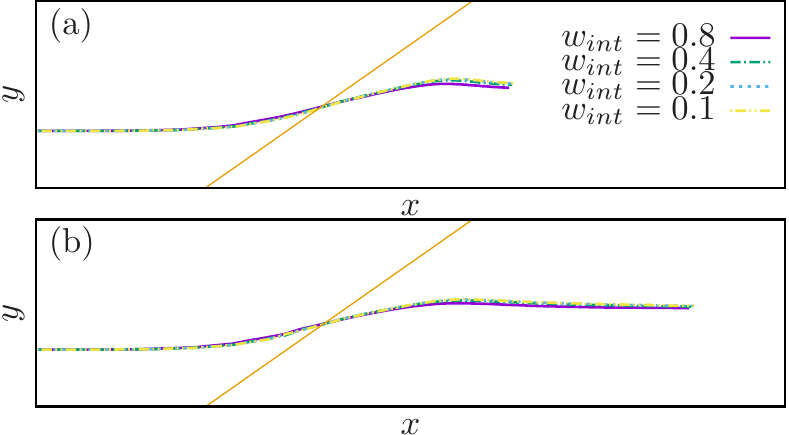}}
    \caption{\label{figthinwinthightheta}(a) Plot of the crack tip trajectories for $\theta=35^o$ and different values of the interface thickness. The load is such that $G/G_c=1.35$. In (b) the same is plotted with  $G/G_c=1.43$. The load values are respectively below and above   the threshold value obtained in the infinitely tough case presented in fig.\ref{plot_inf_th} (b).}
  \end{figure}

 Now  we turn to the case where the kink angle is \textit{large}. More
 specifically we consider values  about 30 degrees. In this case, for all values
 of the load (up to values for which the crack can propagate in the region II)
 the crack does not propagate along the interface and the system always evolves
 toward a situation where a crack has entered (to some extent)   the high
 fracture energy zone. For high  values of $w_{int}$  typical paths are shown in
 fig. (\ref{figthickint}) while for low values a typical path is shown in fig.
 (\ref{figthinwintlowtheta})  In the latter case, a
 closer look at the crack profile together with the boundary between the two
 phases indicates clearly that, as previously, a significant amount of the
 diffuse crack
 is in the region II and therefore participates in  the increase in
 apparent fracture energy and to the unphysical crack behaviour.  
 \begin{figure}
   \centerline{\includegraphics[width=0.6\textwidth]{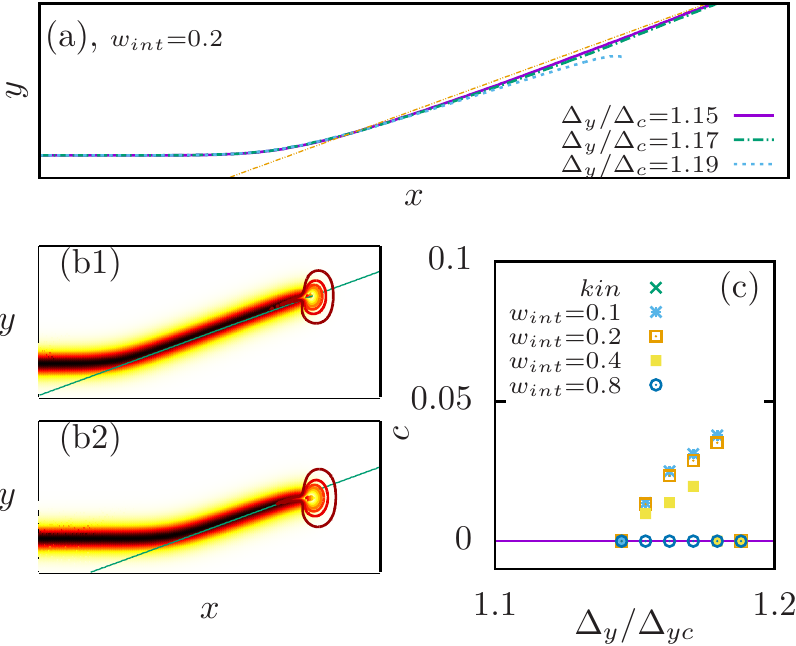}}
 \caption{\label{figthinwintmidtheta} The value of $\theta$ is 20 degrees. (a) Plot of the crack tip trajectories, for three values of the load. For the two lower values, the crack is propagating along the interface while for the higher values (below the propagation threshold in region II) it  goes through the interface and stops.(b1) and (b2), color plots of the phase field, the elastic energy density is plotted using a contour  together with the position of the interface between region I and II. Here $w_{int}=0.1$. In b1 the load is $\Delta_y=1.15\Delta_c$ while in b2 it is  $\Delta_y=1.19\Delta_c$. One can see that in both cases, a significant part of the crack is in the region II where fracture energy is high. (c) crack tip velocities  along the interface as a function of the applied load for different values of the  thickness  of the interface between the two regions. The result from the previous section is also plotted as a reference. One should note that for high values of the load the crack stops for any value of $w_{int}$.}
  \end{figure}

 From this latter set of simulations it is clear that the phase field model used 
 here with the \textit{naive} description of fracture toughness inhomogeneities  cannot
 describe properly crack propagation in an environment where the fracture energy
 varies abruptly. 
 While for lower values of the kink angle the agreement could be
 described as qualitative, here, there is a strong qualitative difference that
 will dramatically affect the transition from interfacial to bulk  cracks. 
 This is due to the diffuse interface modelling of crack that
 implies that when a diffuse crack meets a heterogeneity, it will, regardless
 of the  sharpness of  the boundary between the two regions, \textit{feel} the high energy region.
 Hence it is extremely likely that this result will extend to other approaches,
 including modelling approaches where the relaxational kinetics of the phase
 field used here is replaced by an actual minimisation of the free energy
 functional.

 For the sake of completeness, and to bridge the gap between these two limit
 cases, results obtained for intermediate values of  $\theta$ is briefly
 discussed here. We consider the value  $\theta=20^o$, and present simulation
 results obtained for small values of $w_{int}$. In fig.
 \ref{figthinwintmidtheta}(a) trajectories of the crack for varying values of
 the applied load are presented. Surprisingly, when the applied load is
 increased two successive transitions are observed. At first the crack does not
 propagate as expected, and then  above a load threshold it propagates. Finally
 above a second load threshold that is a function of the interface thickness the
 crack enters the region II and stops. It should be noted that however small is
 $w_{int}$, the second threshold is low as is illustrated in fig.
 \ref{figthinwintmidtheta}(c) where the interfacial crack velocity is plotted
 against the load. Hence in the case of intermediate angles, a completely
 unphysical behaviour is also observed. 
  
 The reason for this is made clear when considering the crack, elastic
 energy density field and obstacle position as in
 fig.\ref{figthinwintmidtheta}~(b1, b2). Indeed, one can see that the crack
 interface is clearly inside the region II, which implies that the fracture
 energy is higher than the one in the region II. The fact that the crack has
 entered the region II is related to the mode mixity that is illustrated by the 
 asymmetric repartition of the elastic energy density with respect to the axis
 of the kinked crack. Hence, the interplay between the diffuse interface
 representation of the crack and a heterogeneous fracture energy field prevents
 the phase field model from properly describing crack propagation when it could
 propagate in the different phases of a multiphasic material.

 \section{Conclusion}

 Here we have considered a simple setup to  characterize how well a
phase field model can describe the propagation of a crack in a heterogeneous
material where the heterogeneities correspond to abrupt changes in the fracture
energy. Simulations show that  in the case of heterogeneities that are
sufficiently tough to prevent crack propagation, a qualitative agreement with
theoretical predictions made using LEFM. Moreover a quantitative comparison
indicates that the small differences can be attributed to a steepening of the
crack profile,

In the case where the fracture energy in the heterogeneity is finite but
significantly larger than in the main region, the interplay between the diffuse
crack and the boundary between the two regions implies significant  changes in
the \textit{apparent} fracture energy that lead to unphysical behaviour. More
specifically the mode mixity along the interfacial crack path implies that the
crack will partially propagate in the high fracture energy region while the
desired  physical behaviour would be an interfacial crack propagating in the low
fracture energy region. {\color{red}This, may  give a rationale to recent
numerical results in more complex systems such as clay composites\cite{msekh2018} where it
has been found that the material strength is decresing with the thickness of the weak interphase between 
hard silicates inclusion and the mattrix}

 The question of whether a not so \textit{naive} description can be
 proposed is open. When the interface plays a significant role because its
 properties differ from bulk properties, alternatives have been proposed
 \cite{Yvonnet,Carolo2018}. They rely on the use of cohesive zone models to
 describe the interface failure and care has been taken to ensure a proper
 transition from bulk cracks to interfacial cracks and the added complexity to
 the plain phase field modelling is simply the translation of the actual
 complexity of the materials. However, in the case of fully bonded systems where
 the interface between the phases  is a simple geometric object, a proper simple
 description of cracks propagation is still missing. Since the unphysical
 behaviour observed here  is strongly related to the diffuse description of the
 crack  an explicit description of the transition from one region to another
 region is likely to be needed.  
 \section*{acknowledgments}
I am grateful to the referees of an earlier version of this manuscript for their
constructive criticisms. I also wish to thank  A. Rowe for carefully checking
the english of this manuscript.

\end{document}